# Generation of a strong reverse shock wave in the interaction of a high-contrast high-intensity femtosecond laser pulse with a silicon target

Kamalesh Jana, Amit D. Lad, Moniruzzaman Shaikh, V. Rakesh Kumar, Deep Sarkar, Yash M. Ved, John Pasley, Alex P. L. Robinson, and G. Ravindra Kumar

# Generation of a strong reverse shock wave in the interaction of a high-contrast high-intensity femtosecond laser pulse with a silicon target

Kamalesh Jana,[1] Amit D. Lad,[1] Moniruzzaman Shaikh,[1] V. Rakesh Kumar,[1] Deep Sarkar,[1] Yash M. Ved,[1] John Pasley,[2] Alex P. L. Robinson,[3] and G. Ravindra Kumar[1,a]

**AFFILIATIONS**

[1]Tata Institute of Fundamental Research, Dr. Homi Bhabha Road, Colaba, Mumbai 400005, India
[2]Department of Physics, York Plasma Institute, University of York, Heslington, York YO10 5DD, United Kingdom
[3]Central Laser Facility, STFC Rutherford Appleton Laboratory, Harwell Campus, Didcot OX11 0QX, United Kingdom

[a]Electronic mail: grk@tifr.res.in

**ABSTRACT**

We present ultrafast pump-probe reflectivity and Doppler spectrometry of a silicon target at relativistic laser intensity. We observe an unexpected rise in reflectivity to a peak approximately ∼9 ps after the main pulse interaction with the target. This occurs after the reflectivity has fallen off from the initially high "plasma-mirror" phase. Simultaneously measured time-dependent Doppler shift data show an increase in the blue shift at the same time. Numerical simulations show that the aforementioned trends in the experimental measurements correspond to a strong shock wave propagating back toward the laser. The relativistic laser-plasma interaction indirectly heats the cool-dense ($n_e \geq 10^{23}$ cm$^{-3}$ and $T_e \sim 10$ eV) target material adjacent to the corona, by hot electron induced return current heating, raising its temperature to around 150 eV and causing it to explode violently. The increase in reflectivity is caused by the transient steepening of the plasma density gradient at the probe critical surface due to this explosive behavior.

Published under license by AIP Publishing. https://doi.org/10.1063/1.5097918

High-intensity, femtosecond laser pulses are capable of producing hot, dense plasma.[1–3] A good understanding of the properties of hot-dense plasma is required to study high energy density science[1] as well as for progress in applications such as plasma based accelerators,[4–7] radiation sources,[8–10] and inertial confinement fusion (ICF) research.[11] The generation and evolution of intense laser-produced plasmas are still not very well understood. The plasma conditions during intense laser-matter interaction change on an ultrafast time scale.[2,12–17] Experimentally tracking the ultrafast dynamics of plasma in the interaction region is a challenging proposition due to the exceptionally high temporal and high spatial resolutions required for this task. Pump-probe reflectometry and Doppler spectrometry are effective tools in understanding the ultrafast dynamics of plasma and thereby evolution of shock waves.[13–18] Previous studies[13–16] on ultrafast lase-plasma dynamics have illuminated a number of aspects of the hydrodynamic nature of plasma in the first few picoseconds of the interaction. It also uncovered a previously unidentified hydrodynamic process[16] in which sound waves are generated by a traffic-jam like instability occurring in the plasma due to the laser prepulse.

These studies[13–16] have highlighted the importance of using diagnostics with a picosecond or even a subpicosecond temporal resolution, also indicating that the evolution in the first 20 ps can sometimes be quite complex. The precise trajectory of the evolution is seen to be sensitive to the laser prepulse levels which essentially determine the nature of the "target" with which the main pulse interacts, and such phenomena can be observed by an optical probe. High levels of prepulses present an extended low-density plasma in which a range of dynamics may be seen,[13,16] and the separation between the pump and probe critical surface can become important. On the other hand, having no prepulse means that the optical probe essentially reflects from the surface of the target plasma at the same location at which the pump is being absorbed.

In the present study, a system is considered in which the laser contrast is high; yet, there is sufficient energy in the prepulse to form a limited preplasma. This turns out to have an interesting consequence as the cold target is now sitting sufficiently close to the probe critical surface. The cold target explodes under the influence of the intense pump pulse, driving a strong shock outward into the preplasma, where it is witnessed by changes in the probe reflectivity and Doppler shift. A silicon target is illuminated by a laser with a relativistically high intensity ($I_L \sim 2 \times 10^{19}$ W/cm$^2$). Time-resolved reflectivity shows a sudden rise and associated increased blue-shift, $\sim$7 ps after the interaction of the main pulse with the target. Targets of intermediate-atomic number (Z) such as silicon are of interest for such ultrafast hydrodynamic studies because the frequency of sound waves that may be supported in a medium is a strong function of Z. It is this fact which enabled the previous observation of 1.9 THz sound waves in glass targets.[16]

The experiment was carried out (setup shown in Fig. 1) at the Tata Institute of Fundamental Research, Mumbai, using a chirped pulse amplification (CPA) based 100 TW Ti:Sapphire laser system which can deliver 800 nm, 25 fs laser pulses at a repetition rate of 10 Hz. A p-polarized 800 nm pump laser pulse was focused by an f/2.5 off-axis parabolic mirror to an $\sim$14 $\mu$m focal spot (FWHM) on a 500 $\mu$m thick optically polished doped silicon ($\langle 100 \rangle$, n-type, phosphorus doped, doping concentration: $10^{17}$ cm$^{-3}$) wafer at an angle of incidence of $\sim$45$°$. The peak intensity was $\sim 2 \times 10^{19}$ W/cm$^2$. The intensity contrast (pedestal to peak ratio) of the laser pulse [measured using a third-order cross-correlator (SEQUOIA)] at 25 ps was $\sim 10^{-9}$. A motorized high-precision stage was used to move the target after each laser shot so that every time the laser interacted with an unblemished region of the target surface. A small fraction of the main laser was extracted by a beam splitter, and it was converted to its second harmonic (400 nm) by a $\beta$-barium borate (BBO) crystal. A BG-39 filter was used after BBO to eliminate the residual 800 nm laser. This second harmonic laser (400 nm) was used as the probe pulse (duration $\sim$80 fs) which reflects from its critical surface ($n_e \sim 7 \times 10^{21}$ cm$^{-3}$) and remains in the overdense region for the pump pulse. The probe was focused on the target to an $\sim$80 $\mu$m spot by an f/12 focusing lens. The probe intensity on the target was $\sim 10^{11}$ W/cm$^2$. Spatial matching between the pump and the probe was confirmed using high resolution imaging. The probe was time delayed with respect to the pump by a high-precision motorized retro-reflector delay stage. The reflected probe was collected by a lens and sent to a photodiode (PD) and a high resolution spectrometer (resolution $\sim$0.03 nm) to measure the reflectivity and Doppler shift simultaneously as a function of time delay with respect to the pump pulse. BG-39 filters were placed in front of the photodiode and spectrometer to avoid unwanted noise.

Figure 2(a) shows the time-resolved probe reflectivity. At a negative time delay (when the probe arrives at the target surface before the pump), the probe gets reflected from a cold silicon target and gives flat, relatively low reflectivity. When the probe arrives simultaneously with the pump pulse, it gets reflected from a plasma mirror created by the pump pulse[19] and there is a sharp rise in the probe reflectivity. The time zero is the time instant at which the reflectivity starts to increase suddenly. After reaching a peak, the reflectivity begins to decrease rapidly to $\sim$20% at $\sim$7 ps. After that, there is a clear rapid rise in the reflectivity up to $\sim$35% (at $\sim$9 ps), followed by a further decay at later times (>10 ps).

We simultaneously measured the Doppler shifts of the probe reflected from its critical layer. The time dependent Doppler shifts of the reflected probe spectra are presented in Fig. 2(b). The Doppler shifts also correlate with reflectivity peaks and dips. For an instant, a rise in the blue shift is observed at the same time (at $\sim$7 ps) when the reflectivity starts to increase. This reflectivity dip matches with a decrease in the blue shift (at $\sim$7 ps), and the second reflectivity peak matches with the enhanced blue shift (at $\sim$9 ps).

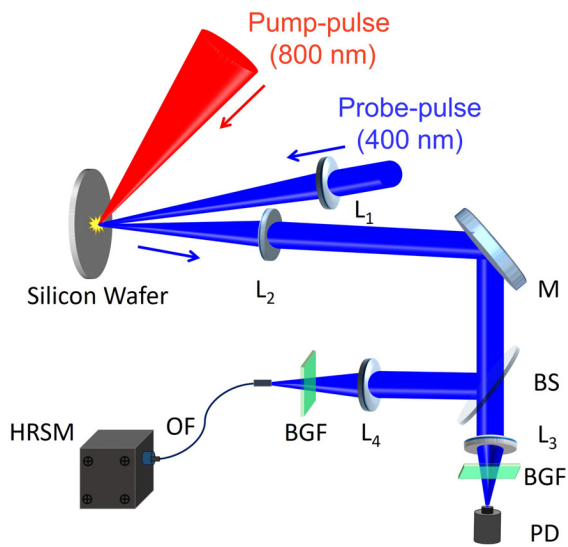

FIG. 1. Schematic of the experimental setup for simultaneous measurements of reflectivity and the Doppler shift of the probe pulse reflected from hot-dense plasma generated on a silicon wafer; M: mirror, $L_1$–$L_4$: lenses, BS: beam splitter, PD: photo-diode, HRSM: high resolution spectrometer, OF: optical fiber, and BGF: blue-green (BG-39) filter.

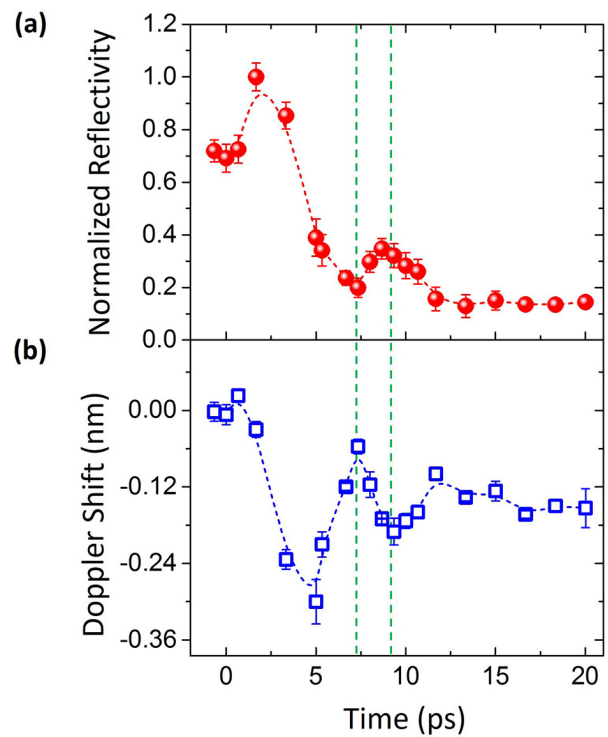

FIG. 2. Simultaneously measured time-resolved (a) reflectivity and (b) Doppler shift of the probe. Anticorrelations between reflectivity and the Doppler shift are indicated by the green vertical dashed lines.

We also performed simulations to understand the experimental observations. Initially, the radiation-hydrodynamics code Hyades was used to model the interaction of the laser prepulse with the silicon target. It uses Sesame Equation-of-State (EOS), a multigroup radiation diffusion transport calculation, to handle thermal radiation and flux-limited diffusion to represent the electron transport. The output of this calculation is used to initialize a Hybrid particle-in-cell (PIC) simulation of the pump interaction. The output of this Hybrid PIC code is then used as the basis for a second Hyades simulation to investigate the hydrodynamics in the first few tens of picoseconds following the interaction of the main pulse. Finally, the output from this second Hyades run is postprocessed to assess the plasma reflectivity as well as various other plasma parameters of the probe critical surface each time to understand the key factors which are determining trends in the reflectivity. Figure 3(a) shows the time-dependent simulated probe reflectivity. The simulated time-dependent reflectivity [Fig. 3(a)] closely resembles that observed in the experiment [Fig. 2(a)]. The initial (at t = 0) difference between the experimental and simulated reflectivity is due to the fact that the reflectivity analysis is done using a collisional plasma model which cannot handle steep gradients in the cold material correctly. Figure 3(b) shows the time-dependent simulated plasma scale length at the probe critical surface, which was found to be the dominant factor in producing the simulated time-dependent reflectivity trend.

The strong peak in reflectivity that occurs at around 10 ps in the simulation is produced by a rapid reduction of the plasma scale length at the critical surface. In turn, this rapid decrease in the plasma scale length at the critical surface appears to be caused by the explosion of the initially cool dense plasma into the low density corona. The cool dense material sitting closest to the corona is rapidly heated by return current setup by the forward streaming hot electrons produced from the relativistic laser-plasma interaction.[20,21] This rapidly reverses the pressure gradient across the density ramp between the corona and the dense plasma, driving a shock wave out into the lower density plasma with which the probe is interacting, resulting in the trends in probe reflectivity and the Doppler shift observed at around 9 ps in the experiment and at around 10 ps in the simulation. A shock wave is also driven inward, into the cold dense solid which is clearly observed in the simulations. This shock wave propagates in the material that is too dense to be accessed by the 400 nm probe and hence not observed here. Figure 4 shows the density profile at 10 ps, obtained from the simulation. As can be seen from Fig. 4, the reverse shock is traveling left and sitting at approximately x = 0 and the forward shock is traveling right and sitting at approximately x = 7.3 $\mu$m. The hot coronal plasma is almost totally ionized. Lower degrees of ionization are present in the cooler, denser regions of the target although mean ionization is always >1 in all the regions of interest. The hot dense region has a $\bar{Z}$ value of 8. Even in the hottest, lowest density regions of the target which are significant (the hot plasma around the critical density surface), the electron-ion mean free paths come out at around 4 nm, meaning that the shock wave regions of the plasma that are of interest are strongly collisional. The simulation also indicates that the reverse shock speed is $\sim 2.7 \times 10^7$ cm/s and the sound speed in the medium through which it is propagating is $\sim 1.04 \times 10^7$ cm/s; thereby, the Mach number of the reverse shock is around 2.6. The simulations show good agreement between the compression ratio and the Mach number in the reverse shock as compared to Eq. (3) in Ref. 22.

Further investigation showed that (a) changing ionization models, even to the point of eliminating ionization, did not significantly change the results and (b) the occurrence of a second peak in the reflectivity curve can be prevented if the temperatures in the dense region are artificially reduced to the point where no reverse shock forms. Hence, the observation of the second peak in the reflectivity curve clearly supports the formation of a reverse shock.

In conclusion, we have presented time-resolved pump-probe reflectivity and Doppler spectrometry of a silicon target at relativistic intensity. Time-resolved reflectivity data show an unexpected rise in reflectivity to a peak approximately ∼9 ps after the interaction of the main laser pulse with the target. Simultaneously, time-dependent Doppler spectrometry shows an enhanced blue shift at the same time. Experimental results well supported by numerical simulations indicate that a strong shock wave is propagating out of the target (toward the laser). This shock wave is formed at the interface between the hot and cool-dense target regions that exist at the end of the prepulse. The cool-dense target material on one side of this interface is rapidly heated by the return current setup by the forward streaming hot electrons produced from the relativistic laser-plasma interaction. The heated cool-dense target material explodes violently into the hot low-density plasma formed by the prepulse and drives a strong shock wave outward into the lower density plasma. The increase in reflectivity is caused by rapid reduction of the plasma scale length at the critical surface, due to this explosive behavior. A forward going shock is also expected to be launched, but this is not observed with these high contrast pulses due to it propagating only in the dense plasma beyond the reach of the 400 nm probe. Our study shows that pump-probe Doppler spectrometry and reflectometry are very effective diagnostics to measure early time rapid plasma dynamics and thereby uncover previously unidentified hydrodynamic phenomena.

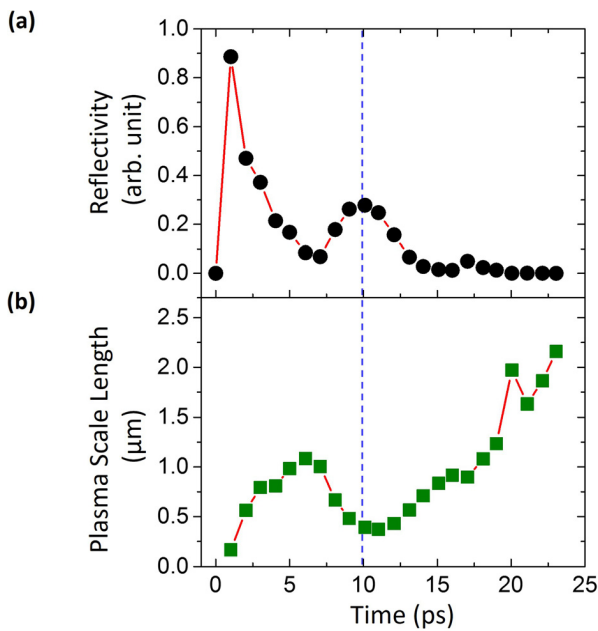

FIG. 3. Time dependent (a) probe reflectivity and (b) plasma scale length at the probe critical surface obtained from numerical simulations. The blue dashed vertical line indicates the time-point where reflectivity peaks and the plasma scale length reduces.

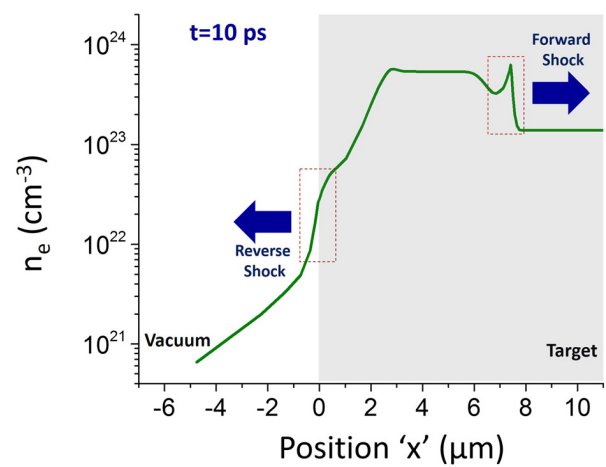

FIG. 4. Density profile at 10 ps obtained from the simulation. The directions of the reverse shock (x $\sim$ 0) and forward shock (x $\sim$ 7.3 $\mu$m) are indicated by blue thick arrows. The laser is incident from the left hand side.

G.R.K. acknowledges the J. C. Bose Fellowship Grant (No. JCB-037/2010) from the Science and Engineering Research Board, Government of India. The authors thank Sudipta L. Roy, Chandra V. Kotyada, and B. Om Subham for their help with the experiment.

## REFERENCES

[1] R. P. Drake, *High Energy Density Physics* (Springer-Verlag, Berlin, Heidelberg, 2006).
[2] P. Gibbon, *Short Pulse Laser Interactions with Matter* (Imperial College Press, 2005).
[3] G. A. Mourou, T. Tajima, and S. V. Bulanov, Rev. Mod. Phys. **78**, 309 (2006).
[4] J. Faure, Y. Glinec, A. Pukhov, S. Kiselev, S. Gordienko, E. Lefebvre, J. P. Rousseau, F. Burgy, and V. Malka, Nature **431**, 541 (2004).
[5] R. A. Snavely, M. H. Key, S. P. Hatchett, T. E. Cowan, M. Roth, T. W. Phillips, M. A. Stoyer, E. A. Henry, T. C. Sangster, M. S. Singh, S. C. Wilks, A. MacKinnon, A. Offenberger, D. M. Pennington, K. Yasuike, A. B. Langdon, B. F. Lasinski, J. Johnson, M. D. Perry, and E. M. Campbell, Phys. Rev. Lett. **85**, 2945 (2000).
[6] H. Chen, S. C. Wilks, J. D. Bonlie, E. P. Liang, J. Myatt, D. F. Price, D. D. Meyerhofer, and P. Beiersdorfer, Phys. Rev. Lett. **102**, 105001 (2009).
[7] L. Robson, P. T. Simpson, R. J. Clarke, K. W. D. Ledingham, F. Lindau, O. Lundh, T. McCanny, P. Mora, D. Neely, C.-G. Wahlstrom, M. Zepf, and P. McKenna, Nat. Phys. **3**, 58 (2007).
[8] H. Schwoerer, P. Gibbon, S. Dusterer, R. Behrens, C. Ziener, C. Reich, and R. Sauerbrey, Phys. Rev. Lett. **86**, 2317 (2001).
[9] Y. T. Li, W.-M. Wang, C. Li, and Z.-M. Sheng, Chin. Phys. B **21**, 095203 (2012).
[10] S. Cipiccia, M. R. Islam, B. Ersfeld, R. P. Shanks, E. Brunetti, G. Vieux, X. Yang, R. C. Issac, S. M. Wiggins, G. H. Welsh, M.-P. Anania, D. Maneuski, R. Montgomery, G. Smith, M. Hoek, D. J. Hamilton, N. R. C. Lemos, D. Symes, P. P. Rajeev, V. O. Shea, J. M. Dias, and D. A. Jaroszynski, Nat. Phys. **7**, 867 (2011).
[11] S. Atzeni and J. Meyer-ter-Vehn, *The Physics of Inertial Fusion: Beam Plasma Interaction, Hydrodynamics, Hot Dense Matter* (Oxford University Press, Oxford, 2004).
[12] P. Gibbon and E. Forster, Plasma Phys. Controlled Fusion **38**, 769 (1996).
[13] S. Mondal, A. D. Lad, S. Ahmed, V. Narayanan, J. Pasley, P. P. Rajeev, A. P. L. Robinson, and G. R. Kumar, Phys. Rev. Lett. **105**, 105002 (2010).
[14] A. Adak, D. R. Blackman, G. Chatterjee, P. K. Singh, A. D. Lad, P. Brijesh, A. P. L. Robinson, J. Pasley, and G. R. Kumar, Phys. Plasmas **21**, 062704 (2014).
[15] K. Jana, D. R. Blackman, M. Shaikh, A. D. Lad, D. Sarkar, I. Dey, A. P. L. Robinson, J. Pasley, and G. R. Kumar, Phys. Plasmas **25**, 013102 (2018).
[16] A. Adak, A. P. L. Robinson, P. K. Singh, G. Chatterjee, A. D. Lad, J. Pasley, and G. R. Kumar, Phys. Rev. Lett. **114**, 115001 (2015).
[17] M. Shaikh, K. Jana, A. D. Lad, I. Dey, S. L. Roy, D. Sarkar, Y. M. Ved, A. P. L. Robinson, J. Pasley, and G. R. Kumar, Phys. Plasmas **25**, 113106 (2018).
[18] T. Ao, G. Chiu, A. Forsman, and A. Ng, AIP Conf. Proc. **505**, 971 (2000).
[19] C. Thaury, F. Quere, J.-P. Geindre, A. Levy, T. Ceccotti, P. Monot, M. Bougeard, F. Reau, P. d'Oliveira, P. Audebert, R. Marjoribanks, and Ph. Martin, Nat. Phys. **3**, 424 (2007).
[20] E. Martinolli, M. Koenig, F. Amiranoff, S. D. Baton, L. Gremillet, J. J. Santos, T. A. Hall, M. Rabec-Le-Gloahec, C. Rousseaux, and D. Batani, Phys. Rev. E **70**, 055402 (2004).
[21] E. Nardi, Z. Zinamon, and I. Maron, Laser Part. Beams **33**, 245 (2015).
[22] W. L. Zhang, B. Qiao, T. W. Huang, X. F. Shen, W. Y. You, X. Q. Yan, S. Z. Wu, C. T. Zhou, and X. T. He, Phys. Plasmas **23**, 073118 (2016).